\documentclass[aps,twocolumn,a4paper,superscriptaddress,nofootinbib]{revtex4-1}
\usepackage{graphicx,subcaption,amsmath,amsfonts,amssymb,multirow,extarrows,bm,acronym,float,caption}
\usepackage[colorlinks,linkcolor=blue,citecolor=blue,urlcolor=blue ]{hyperref}
\floatstyle{plaintop}
\restylefloat{table}
\newcommand{\TRC}{TianQin Research Center for Gravitational Physics and School of Physics and Astronomy, Sun Yat-sen University (Zhuhai Campus), Zhuhai 519082, People's Republic of China}

\newcommand{\PMO}{Key Laboratory of Dark Matter and Space Astronomy, Purple Mountain Observatory, Chinese Academy of Sciences, Nanjing 210023, People's Republic of China}
\newcommand{\USTC}{School of Astronomy and Space Science, University of Science and Technology of China, Hefei, Anhui 230026, People's Republic of China}
\newcommand{\SCUT}{School of Physics and Optoelectronics, South China University of Technology, Guangzhou 510641, People's Republic of China}

\begin{document}

\title{Probing chromatic onsets of gravitational wave overtones}

\author{Hai-Tian Wang}
\affiliation{\PMO}
\affiliation{\USTC}
\affiliation{\TRC}
\author{Yi-Ming Hu}
\affiliation{\TRC}
\author{Peng-Cheng Li}
\affiliation{\SCUT}
\author{Yi-Zhong Fan}
\email{yzfan@pmo.ac.cn}
\affiliation{\PMO}
\affiliation{\USTC}

\date{\today}

\begin{abstract}
The ringdown data of both GW150914 and GW190521\_074359 (GW190521r) show evidence supporting the presence of overtone. 
Previous studies all adopt a fundamental assumption, which were motivated more by convenience than by first principles, that the first overtone and the fundamental mode share a same onset. 
In this work, for the first time we relax such assumption, and we aim to probe the possible chromatic onsets of these two components within the GW150914 and GW190521r ringdown data. 
For both events, we bound the onset lags to be $\Delta t_0\geq 5M_f$ at probabilities of $\geq 94.7\%$, where $M_f$ is the mass of the remnant black hole formed in the merger.  
This result moderately favours the non-simultaneous excitation between the fundamental mode and the first overtone in the ringdown. 
\end{abstract}

\maketitle

\acrodef{GW}{gravitational wave}
\acrodef{LIGO}{Laser Interferometer Gravitational-Wave Observatory}
\acrodef{LVC}{LIGO-Virgo Collaboration}
\acrodef{BNS}{binary neutron star}
\acrodef{NS}{neutron star}
\acrodef{BH}{black hole}
\acrodef{BBH}{binary black hole}
\acrodef{NSBH}{neutron star-black hole}
\acrodef{IMBH}{intermediate mass black hole}
\acrodef{GR}{general relativity}
\acrodef{PN}{post-Newtonian}
\acrodef{SNR}{signal-to-noise ratio}
\acrodef{PSD}{power spectral density}
\acrodef{PDF}{probability density function}
\acrodef{FIM}{fisher information matrix}
\acrodef{ppE}{parametrized post-Einsteinian}
\acrodef{IMR}{inspiral-merger-ringdown}
\acrodef{QNMs}{quasinormal modes}
\acrodef{ISCO}{innermost-stable circular orbit}

\section{Introduction}\label{sec:intro}

A single distorted black hole will be formed after the violent plunge of compact binary black holes (BBHs). 
The gravitational wave (GW) radiation in this stage is called ringdown, after which it settles to a stationary state. 
The ringdown stage is described by the superposition of quasinormal modes (QNMs), in the form of damped sinusoids \citep{Schw_PRD_Vishveshwara1970, GW_APJL_Press1971, QNM_APJ_Teukolsky1973, 1999CQGra..16R.159N}. 
The QNMs can be decomposed into spin-weighted spheroidal harmonics with angular indices $(l,m)$, each consists of a set of overtones $n$. 
Recently, \citet{NoHair_PRL_Isi2019} find evidence for the existence of the first overtone mode $(n=1)$ in the ringdown signal of GW150914 \citep{LIGO_PRL_GW150914}. 
Furthermore, \citet{Overtone_PRX_Giesler2019} confirm the importance of the overtone modes by analyzing ringdown signal from numerical relativity (NR) simulation. 
Later on, \citet{LIGOScientific:2020tif} reveal the existence of the first overtone mode in GW190521\_074359 (GW190521r). 

Normally, analyses on ringdown data are performed in the time domain (TD) due to the abrupt start \citep{PRD_Carullo2019, NoHair_PRL_Isi2019}. 
It has been developed to extract the properties of the remnant BH with GW data \citep{PRD_Carullo2019, Overtone_PRX_Giesler2019, LIGOScientific:2020tif}. 
TD analysis based on ringdown data has also been used to test no-hair theorems \citep{Hawking:1971vc, PhysRevLett.34.905, NoHair_PRL_Isi2019} and the BH area law \citep{2021PhRvL.127a1103I}. 
A detailed introduction to different TD methods can be found in \citet{2021arXiv210705609I}. 

The start time of the ringdown signal is an important and attractive problem, which is not finally solved yet. 
The numerical relativity (NR) simulations suggest that the ringdown signal can be modeled accurately with the fundamental tone at $10-16 M_f$ after the peak strain amplitude \citep{2012PhRvD..85b4018K, 2014PhRvD..90l4032L, Thrane:2017lqn, 2018PhRvD..97j4065B}, where $M_f$ is the remnant mass. 
This has been supported by the subsequent analyses of the real GW data \citep{PRD_Carullo2019, NoHair_PRL_Isi2019, LIGOScientific:2020tif} as well. 
Moreover, it is found out that the ringdown signal starts from the peak strain amplitude when one takes into account the overtone modes,  supposing that the first overtone and the fundamental mode are present simultaneously from the fit time onward \citep{NoHair_PRL_Isi2019, Overtone_PRX_Giesler2019}.
Though such progresses are intriguing, the possibility that different overtones are not excited simultaneously has not been excluded.  
Hence, a further probe is still needed, which is the main goal of this work. 

Before doing that, let us look into another interesting question:
Do different $(l,m)$ modes have different start times in the ringdown signal? 
Some studies based on waveforms from NR simulations show that different $(l,m)$ modes reach the peak strain amplitude at different times \citep{2008PhRvD..77d4031S, 2011PhRvD..84l4052P, 2018PhRvD..98h4038B}. 
The underlying physical reasons for the peak time shifts among different $(l,m)$ modes around the transition from merger to ringdown are still unclear. 
According to recent studies \citep[i.e.,][]{NoHair_PRL_Isi2019, Overtone_PRX_Giesler2019}, the ringdown signal starts from the peak strain amplitude. 
These chromatic peaks lead to time and phase shifts of QNMs when one matches them to the NR waveforms, implying that different $(l,m)$ modes in QNMs are not excited simultaneously. 
It is therefore reasonable to speculate that different overtones in each $(l,m)$ may have different start times \citep[see also][for theoretical/independent argument]{2020PhRvD.101d4033B}. 

Motivated by the above consideration, in this work we re-analyze the ringdown data of GW150914 and GW190521r, allowing different start times for the first overtone and the fundamental mode. 
For both events, the onset lags between the first overtone and the fundamental mode are bounded to be larger than $5M_f$ at probabilities of $\geq 94.7\%$. 
The method we adopted is introduced in Sec.~\ref{sec:method}, and the main results will be presented in Sec.~\ref{sec:result} and Sec.~\ref{sec:inject}. 
Finally, we summarize this work in Sec.~\ref{sec:summary}. 
We assume $G = c = 1$ throughout the paper, unless otherwise specified. 

\section{Method}\label{sec:method}

In general relativity, the ringdown waveform of Kerr BH is fully described by two time-dependent polarization $h(t)=h_{+}(t)-ih_{\times}(t)$, which can be written as 
\begin{equation}
\begin{aligned}
&h_{+}(t)-ih_{\times}(t) \\
=&\sum_l^{}\sum_m^{}\sum_n^{N}A_{lmn}w(t,\Delta t_n)\exp\left(i(-\Omega_{lmn}(t)\right.\\
&\left.+\phi_{lmn})\right)\times{}_{-2}Y_{lm}(\iota),
\end{aligned}
\label{eq:ringdown}
\end{equation}
where $N$ is the total overtone numbers, $A_{lmn}$ ($\phi_{lmn}$) characterize the amplitudes (phases) of each ringdown mode at the starting time, $\iota$ is the inclination angle, and ${}_{-2}Y_{lm}(\iota)$ are the spin-weighted spherical harmonics \citep{2004CQGra..21..787D, 2018PhRvD..97d4048B, 2014PhRvD..90l4032L, 2021PhRvD.103h4048F}. 
We adopt a cosine-tapered function $w(t)$ as a smooth step function, to preserve the differentiability of the time domain waveform. 
\begin{equation}
w(t,\Delta t_n)=
\left\{\begin{array}{lc} 
\frac{1}{2}\left[1-\cos\left(\frac{\pi t}{\Delta t_n}\right)\right] & 0\leq t<\Delta t_n,\\
1 & \Delta t_n\leq t\leq T.  
\end{array}\right.
\label{eq:wf}
\end{equation}
$T$ is the duration of the ringdown signal. $\Delta t_n$ determine the cosine-tapered regions at the beginning of the ringdown signal, representing the time delay of mode $n$. 
\footnote{If we use a Heaviside window function, the derivative of the strain $dh/dt$ provides the luminosity radiated (time derivative of the energy) of the source and a discontinuity on this magnitude would involve a divergence on the luminosity i.e. on the energy radiated as well. 
We are grateful to the anonymous referee for pointing out that this could be fixed by adding a transference function $w(t)$ that smooths out such divergences. }

The response of a single detector $k$ to the gravitational waves is described as 
\begin{equation}
h_{k}(t)=F_{k}^{+}(\theta, \phi, \psi) h_{+}(t)
+F_{k}^{\times}(\theta, \phi, \psi) h_{\times}(t),
\end{equation}
where $F_k^{+,\times}(\theta, \phi, \psi)$ are the antenna beam patterns, $\theta, \phi$ are the longitudinal and azimuthal angles, and $\psi$ is the polarization angle. 

The gravitational wave data stream $d$ is provided by the Gravitational-Wave Open Science Center \citep{LIGO_PRX2019}, which contains the signal $h(t)$ and noise $n(t)$. 
In standard gravitational wave data analysis, the detector noise can be assumed to be a Gaussian stochastic process \citep{gw150914_PRL2016}. 
As a Gaussian stochastic process, each set $[n(t_0),n(t_1),,,n(t_{N_s-1})]$ is distributed as multivariate Gaussian \ac{PDF}, 
\begin{equation}
\mathbf{n(t)} \sim \mathcal{N}(\boldsymbol{\mu}, \mathbf{\Sigma}),
\end{equation}
where $\boldsymbol{\mu}$ and $\mathbf{\Sigma}_{ij}=\rho(i-j)$ are the mean and the covariance matrix of the noise time series, respectively. 
After applying a high-pass filter with a roll-on frequency of $20$ Hz, the data stream can be treated as zero mean. 
Then we assume that the noise in the detector is stationary. 
$\rho$ is the auto-covariance function 
\begin{equation}
\rho(i-j)=\left\langle n_{i} n_{j}\right\rangle .
\end{equation}

On the one hand, one can get the covariance matrix by calculating the auto-covariance function (ACF) from GW data directly. 
On the other hand, according to the Wiener-Khinchin theorem, the auto-covariance function is the inverse Fourier transform of the \ac{PSD}. 
In this case, the covariance matrix is a circular Toeplitz matrix, then this analysis will be equivalent to an un-windowed Fourier-domain analysis. 
Thus, one should truncate the auto-covariance function with proper duration to break circularity \citep{2021arXiv210705609I}, if it is obtained from the inverse Fourier transform of the PSD. 
We apply these two solutions to GW150194 and GW190521r respectively. 

For gravitational wave data analysis in time domain, the inner product between two waveforms $h_1(t)$ and $h_2(t)$ can be defined as 
\begin{equation}
(h_1 \mid h_2)=h_1^T\Sigma^{-1}h_2.
\label{eq:inner_p}
\end{equation}
Given the observed strain series $d_k(t)$ and the gravitational wave signal $h_k(t)$ from waveform model, the log-likelihood function can be written as 
\begin{equation}
\log \mathcal{L}_k=-\frac{1}{2}(d_k-h_k \mid d_k-h_k) .
\label{eq:logl}
\end{equation}
The log-likelihood function of multiple detectors is the sum of the individual log-likelihood. 

We carry out Bayesian inference with {\sc Bilby} package \citep{Ashton_APJ2019} and {\sc Pymultinest} sampler \citep{Buchner_AAP2014, 2009MNRAS.398.1601F, 2019OJAp....2E..10F} with $5000$ live points. 
For the TD ringdown analyses, Bayesian inferences are performed with $0.5$ seconds data stream that start from the peak strain amplitude. 
The priors on the detector frame final mass $M_f$ are $[50,\, 100]M_{\odot}$ and $[50,\, 120]M_{\odot}$ for GW150914 and GW190521r, respectively. 
Following \citet{NoHair_PRL_Isi2019} and \citet{LIGOScientific:2020tif}, we fix some extrinsic parameters for GW150914 (GW190521r): the geocentric time is set to be $1126259462.408665~(1242459857.472)$ GPS, the right ascension is $\alpha=1.95~(5.46)$ rad, the declination is $\delta=-1.27~(0.59)$ rad, the polarization angle is $\psi=0.82~(1.37)$ rad, and the inclination angle is $\iota=\pi~(2.64)$ rad. 
For other parameters, they are the same for both GW150914 and GW190521r. 
The priors on the dimensionless spin $\chi_f$, the amplitude parameter $A_{lmn}$, and the phase parameter $\phi_{lmn}$ are uniformly distributed in the ranges of $[0,\,0.99]$, $[0,\,25\times 10^{-20}]$, and $[0,\,2\pi]$, respectively. 

\section{Results of $\rm \bf GW150914$ and $\rm \bf GW190521r$}\label{sec:result}

There are $50$ GW events reported by Advanced LIGO and Advanced Virgo during the first two Gravitational-Wave Transient Catalogs \citep{LIGO_PRX2019, LIGO_O3a_PRX2020}. 
Among them, only GW150914 and GW190521r show tentative evidence for the first overtone \citep{LIGOScientific:2020tif}. 
Other events can hardly be adopted in this work since the fundamental mode tends to match the peak of the strain if one does not fix the start time. 
Thus, our analyses focus on GW150914 and GW190521r. 
We assume that the first overtone starts from the peak and then varies the start time of the fundamental mode. 
Equivalently, we set $\Delta t_1=0$ and assume a uniform prior on $\Delta t_0$ in the range of $[0,\, 20]\,M_f$. 
It means that we zero out the fundamental mode before its assigned start time for part of the analysis segment. 
Note that in our analysis the fundamental mode is assumed to start at a time earlier than $20\,M_f$ after the peak, which is well motivated by the previous finding that the fundamental mode already dominates the ringdown signal in $10-20\,M_f$ \citep{2012PhRvD..85b4018K, 2014PhRvD..90l4032L, Thrane:2017lqn,2018PhRvD..97j4065B,NoHair_PRL_Isi2019}.  
Moreover, a $\Delta t_0$ larger than $20M_f$ means that the fundamental mode starts later, which would lead to a lower signal-to-noise ratio and hence the less tight constraints on the physical parameters. 

\begin{figure}
\centering
\includegraphics[width=0.95\linewidth]{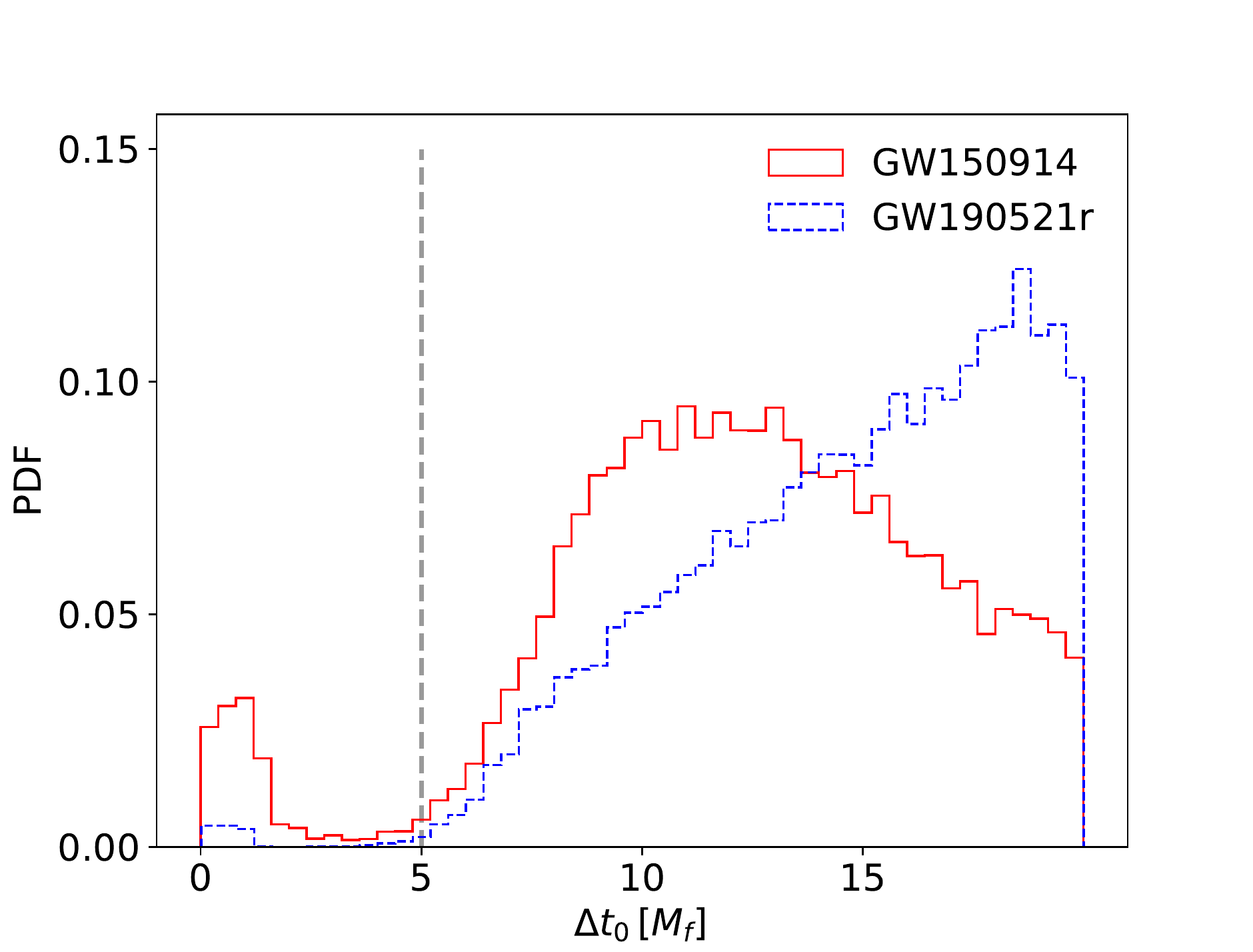}
\caption{
Distributions of $\Delta t_0$ of GW150914 (red histogram) and GW190521r (dashed blue histogram). 
We measure $\Delta t_0=12.38^{+6.50}_{-8.19}~(15.18^{+4.34}_{-7.33})$ at $90\%$ credibility for GW150914 (GW190521r). 
The dashed gray line marks the time $5\,M_f$ later than the peak of the ringdown signal. 
For GW150914 (GW190521r), the probability that $\Delta t_0\geq 5\,M_f$ is $94.7\%$ ($99.3\%$). 
}
\label{fig:dt0s}
\end{figure}

In FIG.~\ref{fig:dt0s}, we summarize the main results of our analysis. 
For both GW150914 and GW190521r, the distributions of $\Delta t_0$ show moderate evidence that there is a deviation between the start times of the overtone mode and the fundamental mode.  
The probability that $\Delta t_0\geq 5\,M_f$ is $94.7\%$ ($99.3\%$) for GW150914 (GW190521r). 
The distributions of $\Delta t_0$ are wide, which range from $5\,M_f$ to $20\,M_f$ and can be explained in two ways. 
On the one hand, the fundamental mode begins to dominate around $10\,M_f$ after the peak strain amplitude. 
On the other hand, we introduce a cosine-tapered smooth function in Eq.~\ref{eq:ringdown} to avoid the discontinuity on the derivative of the ringdown signal. 
This also leads to an ambiguous start time of the fundamental mode in our case. 

\begin{figure*}
\centering
\begin{subfigure}[b]{0.5\linewidth}
\centering
\includegraphics[width=\textwidth,height=8cm]{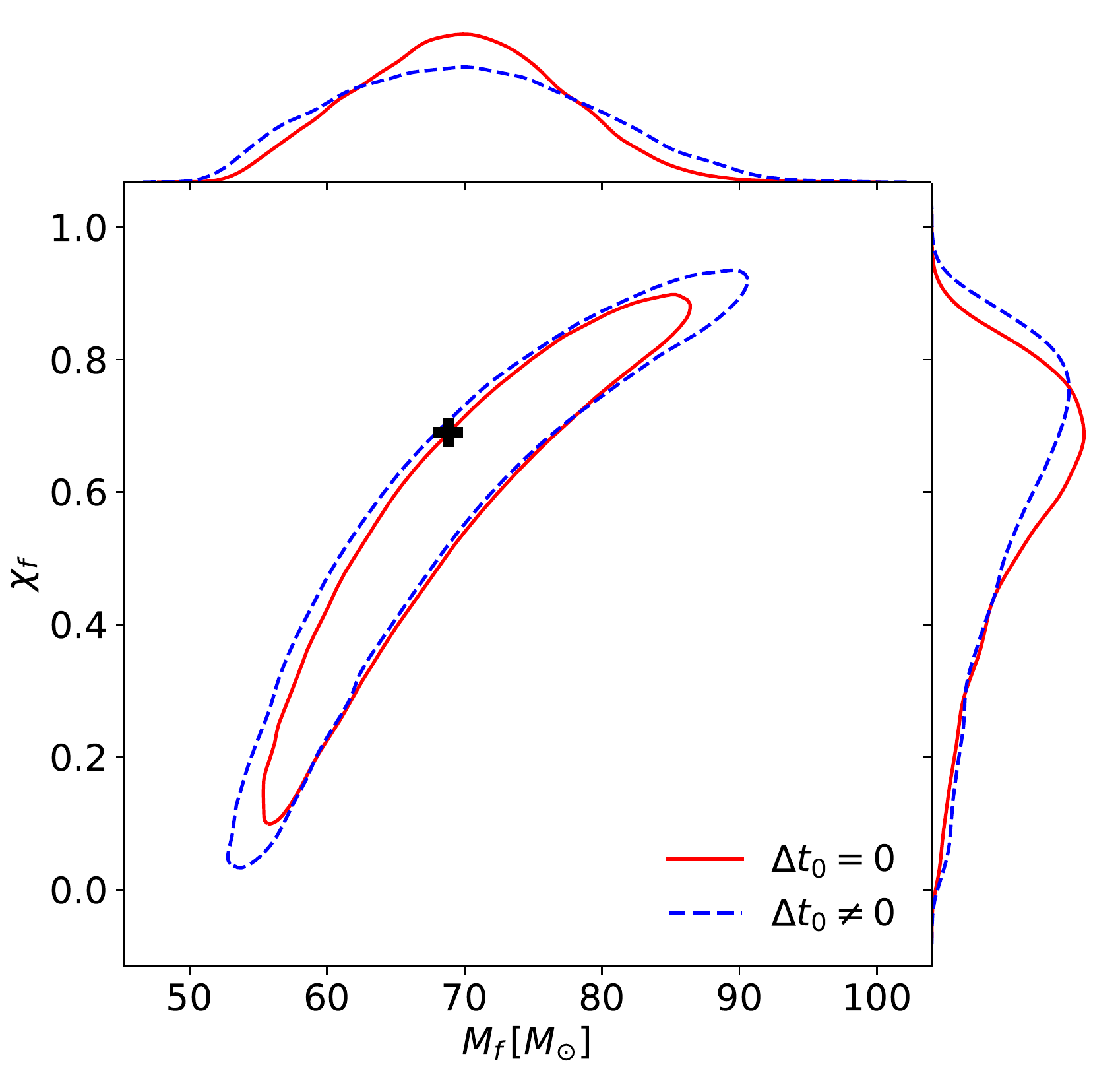}
\end{subfigure}%
\begin{subfigure}[b]{0.5\linewidth}
\centering
\includegraphics[width=\textwidth,height=8cm]{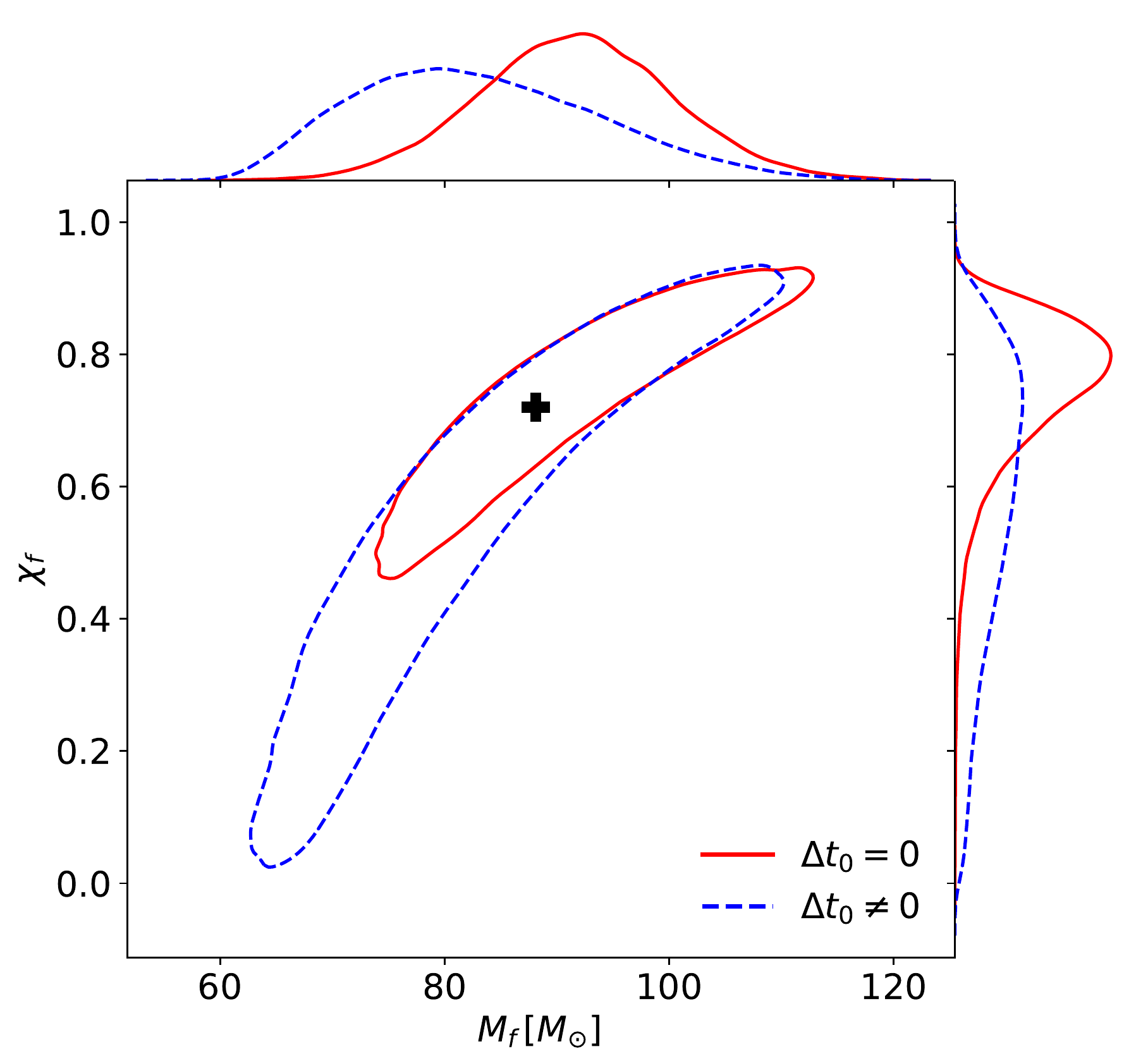}
\end{subfigure}%
\caption{
Posterior distributions of the redshifted final mass and final spin for GW150914 (left panel) and GW190521r (right panel). 
The red (dashed blue) contour represents the result assuming there is a deviation (no deviation) between the overtone mode and the fundamental mode. 
All of these contours are $90\%$-credible regions. 
The 1D marginalized posterior distributions for the final mass and the final spin are shown in the top and right-hand panels, respectively. 
For GW150914 (GW190521r), the median values of the final mass $M_f=68.8\,M_{\odot}~(88.1\,M_{\odot})$ and final spin $\chi_f=0.69\,~(0.72)$ given by the full IMR analysis \citep{LIGOScientific:2020tif}, which is shown as a black plus marker. 
}
\label{fig:mfsf}
\end{figure*}

Additionally, we also perform TD Bayesian inferences on GW150914 and GW190521r with $\Delta t_0=0$, as comparison. 
We show the posterior distributions of the redshifted final masses and spins for GW150914 and GW190521r in FIG.~\ref{fig:mfsf}. 
In the case of $\Delta t_0=0$, our results are in agreement with those of \citet{LIGOScientific:2020tif}, as anticipated. 
Compared with the case of $\Delta t_0=0$, the distributions of the parameters for $\Delta t_0\neq 0$ of both events are broader. 
This is reasonable since an additional parameter ($\Delta t_0$) has been introduced into the ringdown waveform modeling. 

\begin{table}
    \begin{center}
    \setlength{\tabcolsep}{3pt} 
    \renewcommand{\arraystretch}{1.7} 
        \begin{tabular}{@{\extracolsep{3pt}}
        c|c|c|c|c@{}}
            \hline
            \hline
            Events & $M_f[M_{\odot}]$ & $\chi_f$ & $\Delta t_0$ & $\log_{10}\mathcal{B}^{\Delta t_0\neq 0}_{\Delta t_0=0}$  \\
            \hline
            \multicolumn{1}{c|}{\multirow{2}{*}{GW150914}}&
            $69.7^{+12.0}_{-11.7}$ & $0.63^{+0.20}_{-0.41}$ & $0$ &- \\
            \multicolumn{1}{c|}{}&
            $69.8^{+15.0}_{-13.5}$ & $0.64^{+0.22}_{-0.47}$ & $12.38^{+6.50}_{-8.19}$ & $+0.3$ \\
            \hline
            \multicolumn{1}{c|}{\multirow{3}{*}{GW190521r}}&
            $91.6^{+14.3}_{-14.3}$ & $0.77^{+0.12}_{-0.24}$ & $0$ &- \\
            \multicolumn{1}{c|}{}&
            $82.2^{+20.0}_{-14.9}$ & $0.60^{+0.26}_{-0.45}$ & $15.18^{+4.34}_{-7.33}$ & $+0.9$ \\
            \hline
            \hline
        \end{tabular}
    \end{center}
\caption{
The median and symmetric $90\%$-credible intervals, of the redshifted final mass, final spin, and $\Delta t_0$, inferred from the ringdown analyses of GW150194 and GW190521r. 
We quantify the choice of different priors on $\Delta t_0$ using log Bayes factor $\log_{10}\mathcal{B}_{\Delta t_0\neq 0}^{\Delta t_0=0}$. 
The first overtone has been taken into account in the ringdown analyses. 
}
\label{table:compare}
\end{table}

In TABLE.~\ref{table:compare}, we show the corresponding $90\%$-credible measurements of $M_f$, $\chi_f$, and $\Delta t_0$. 
To quantify the contribution of $\Delta t_0$, we calculate the log Bayes factor versus the fixed case. 
For GW150914 (GW190521r), the log Bayes factor of $\Delta t_0\neq 0$ is $\log_{10}\mathcal{B}^{\Delta t_0\neq 0}_{\Delta t_0=0}=+0.3\,(+0.9)$, which moderately indicates the presence of $\Delta t_0$ for both GW150914 and GW190521r. 

\section{Injection test}\label{sec:inject}
The Bayes factors of both GW150914 and GW190521r are too low to reliably suggest a deviation between the overtone mode and the fundamental mode. 
To draw a more robust conclusion, we perform an injection test with pure Gaussian noise and repeat the analysis. 
SXS:BBH:0305 is a specific example the GW150914-like NR waveform in the Simulating eXtreme Spacetimes (SXS) catalog \citep{2019CQGra..36s5006B}. 
This waveform describes a source with a mass ratio of $0.82$ and a remnant with a spin of $\chi_f=0.69$. 
We assume that the chirp mass is $31M_{\odot}$, then the final mass of the remnant is $M_f=68.2M_{\odot}$. 
We assume that the luminosity distance is $450\rm Mpc$, the inclination angle is $\pi/6$, and the sky location is the same as GW150914. 
We inject the $l=m=2$ spheroidal harmonic of this waveform into Gaussian noise that is estimated from GW data around GW150914. 
The network signal-to-noise ratio of the post-peak of this signal is about $14$ when it is detected by both LIGO Hanford and Livingston observatories. 
In this case, similar to GW150914, we only need to consider the first overtone and the fundamental mode in ringdown signal analysis.  

\begin{figure}
\centering
\includegraphics[width=0.95\linewidth]{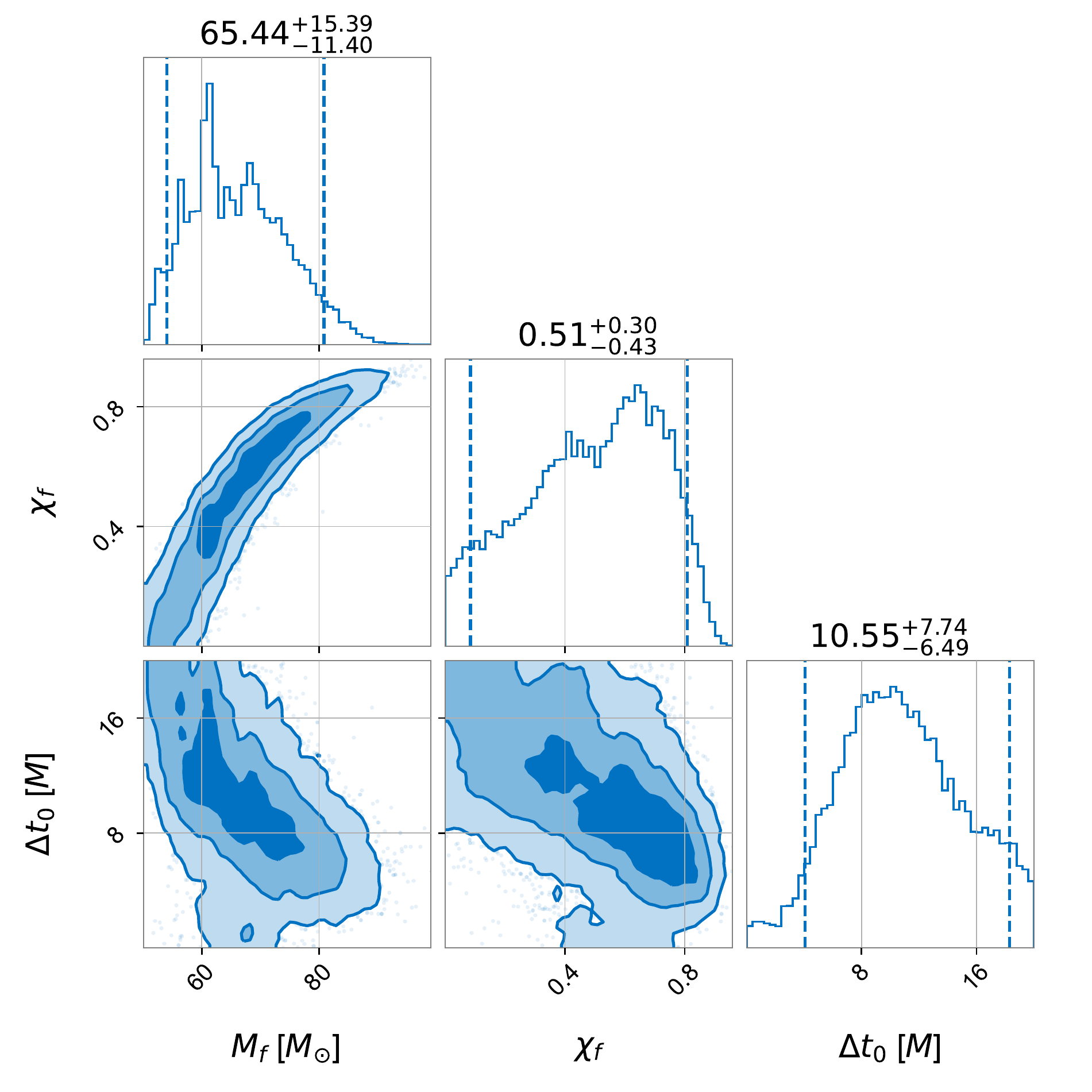}
\caption{
Posterior distributions of the final mass $M_f$, the final spin $\chi_f$, and the deviation between the fundamental mode and the first overtone $\Delta t_0$ in the injection case. 
The contours represent the credible level at $90\%$. 
The numbers listed in the diagonal are the median values and the $90\%$-credible measurements of each parameter.
}
\label{fig:inject_0305}
\end{figure}

Then we perform the time-domain Bayesian inference with similar priors introduced in Sec.~\ref{sec:method}. 
We show posterior distributions of the final mass $M_f$, the final spin $\chi_f$, and the deviation $\Delta t_0$ in Fig.~\ref{fig:inject_0305}. 
Among them, the deviation is constrained to be $\Delta t_0=10.6^{+7.7}_{-6.5}\,M_f$ at $90\%$ credibility, which is similar to the result of GW150914 (see Sec.~\ref{sec:result}). 
Thus, we conclude that the moderate evidence of the deviation between the fundamental mode and the first overtone presents in both the real gravitational wave data and the injection test. 

\section{Conclusions}\label{sec:summary}

In this work, we analyze the ringdown signal of GW150914 and GW190521r with the first overtone and the fundamental mode, allowing different start times of these two components by introducing a cosine-tapered smooth function on the fundamental mode. 
The traditional treatment that assumes a same onset time for different modes is simpler, but lacks the first-principle reasoning to support it; the different onset time assumption, on the other hand, may be more natural and has not been probed before.
Indeed, if we abandon the achromatic onset assumption, for GW150914 and GW190521r 
we have $\Delta t_0\geq 5 M_f$ at the confidence levels of $94.7\%$ and $99.3\%$, respectively. 
To further test our results, we perform a similar analysis on a GW150914-like NR waveform SXS:BBH:0305 and get similar result. 
Anyhow, the Bayes factors for the chromatic onsets of different modes in GW150914 and GW190521r are still low and more high-quality observations are required to reliably probe the onsets of different modes. 


The previous NR simulation results suggest that the ringdown signal may start from $10-16 M_f$ later than the peak strain amplitude, at which the linear perturbation theory works \citep{2014PhRvD..90l4032L,2018PhRvD..97j4065B}. 
However, under the assumption that the fundamental and overtone modes present at the same time, it has been found that the linear description can be extended to the full waveform following the peak of the strain and times around the peak are dominated by ringdown overtones \citep{Overtone_PRX_Giesler2019}. 
On the other hand, although \citet{NoHair_PRL_Isi2019} and  \citet{Overtone_PRX_Giesler2019} find that, when overtones are included, one can obtain a better match that traces the data all the way back to the peak, they do not rule out the possibility that different overtones are not excited simultaneously. 
Interestingly, our analyses suggest that the overtones may be not excited simultaneously \citep[see also][for the theoretical argument]{2020PhRvD.101d4033B}.
Notice, however, that the linearity of the system around the peak is still under debate, for example \citet{2018PhRvD..97j4065B} find the presence of nonlinearity in a sphere of radius about $5M_f$ around the final remnant BH of a BBH spacetime, which means that this region can not be fully described by the perturbation theory. 
This may lead to an earlier onset time of the fundamental mode. 
We aim to relax such assumptions whence better-quality data is available.  

In summary, by abandoning the simultaneous onset assumption for different modes, the ringdown data of GW150914 and GW190521r indicate a moderate chromatic onset of the first overtone and the fundamental mode. 
Moreover, we measure the separation between the start times of the first overtone and the fundamental mode to be lager than $5\,M_f$ at $94.7\%$ credibility. 
Furthermore, our results provides another perspective to understand the time shifts between different $(l,m)$ modes \citep{2008PhRvD..77d4031S, 2011PhRvD..84l4052P, 2018PhRvD..98h4038B}. 
Finally, we stress that the current data can not draw a decisive conclusion and more accumulation of data is needed to better resolve this issue. 

\begin{acknowledgments}
We are grateful to the anonymous referee for the insightful suggestions. 
We also thank Maximiliano Isi and Shao-Peng Tang for the relevant discussions. 
This work has been supported by NSFC under Grants No. 11921003, No. 12173104, and No. 12047550. 
Project also supported by MOE Key Laboratory of TianQin Project, Sun Yat-sen University. 
This research has made use of data, software, and/or web tools obtained from the Gravitational Wave Open Science Center (https://www.gw-openscience.org), a service of LIGO Laboratory, the LIGO Scientific Collaboration, and the Virgo Collaboration. 
LIGO is funded by the U.S. National Science Foundation. 
Virgo is funded by the French Centre National de Recherche Scientifique (CNRS), the Italian Istituto Nazionale della Fisica Nucleare (INFN), and the Dutch Nikhef, with contributions by Polish and Hungarian institutes.

\end{acknowledgments}

\bibliographystyle{aasjournal}
\bibliography{qnm_composition}

\begin{thebibliography}{}
\expandafter\ifx\csname natexlab\endcsname\relax\def\natexlab#1{#1}\fi
\providecommand{\url}[1]{\href{#1}{#1}}
\providecommand{\dodoi}[1]{doi:~\href{http://doi.org/#1}{\nolinkurl{#1}}}
\providecommand{\doeprint}[1]{\href{http://ascl.net/#1}{\nolinkurl{http://ascl.net/#1}}}
\providecommand{\doarXiv}[1]{\href{https://arxiv.org/abs/#1}{\nolinkurl{https://arxiv.org/abs/#1}}}

\bibitem[{Abbott {et~al.}(2016)}]{gw150914_PRL2016}
Abbott, B.~P., {et~al.} 2016, Phys. Rev. Lett., 116, 061102,
  \dodoi{10.1103/PhysRevLett.116.061102}

\bibitem[{Abbott {et~al.}(2019)}]{LIGO_PRX2019}
---. 2019, Phys. Rev. X, 9, 031040, \dodoi{10.1103/PhysRevX.9.031040}

\bibitem[{{Abbott} {et~al.}(2016){Abbott}, {Abbott}, {Abraham}, {Acernese},
  {Ackley}, {Adams}, {Adams}, {Adhikari}, {Adya}, {Affeldt}, \&
  et~al.}]{LIGO_PRL_GW150914}
{Abbott}, R., {Abbott}, T.~D., {Abraham}, S., {et~al.} 2016, \prl, 116, 061102,
  \dodoi{10.1103/PhysRevLett.116.061102}

\bibitem[{Abbott {et~al.}(2021)}]{LIGOScientific:2020tif}
Abbott, R., {et~al.} 2021, Phys. Rev. D, 103, 122002,
  \dodoi{10.1103/PhysRevD.103.122002}

\bibitem[{{Abbott} {et~al.}(2021){Abbott}, {Abbott}, {Abraham}, {Acernese},
  {Ackley}, {Adams}, {Adams}, {Adhikari}, {Adya}, {Affeldt}, \&
  et~al.}]{LIGO_O3a_PRX2020}
{Abbott}, R., {Abbott}, T.~D., {Abraham}, S., {et~al.} 2021, Physical Review X,
  11, 021053, \dodoi{10.1103/PhysRevX.11.021053}

\bibitem[{Ashton {et~al.}(2019)Ashton, Hübner, Lasky, Talbot, Ackley,
  Biscoveanu, Chu, Divakarla, Easter, Goncharov, Vivanco, Harms, Lower,
  Meadors, Melchor, Payne, Pitkin, Powell, Sarin, Smith, \&
  Thrane}]{Ashton_APJ2019}
Ashton, G., Hübner, M., Lasky, P.~D., {et~al.} 2019, The Astrophysical Journal
  Supplement Series, 241, 27, \dodoi{10.3847/1538-4365/ab06fc}

\bibitem[{{Baibhav} {et~al.}(2018){Baibhav}, {Berti}, {Cardoso}, \&
  {Khanna}}]{2018PhRvD..97d4048B}
{Baibhav}, V., {Berti}, E., {Cardoso}, V., \& {Khanna}, G. 2018, \prd, 97,
  044048, \dodoi{10.1103/PhysRevD.97.044048}

\bibitem[{{Bhagwat} {et~al.}(2020){Bhagwat}, {Forteza}, {Pani}, \&
  {Ferrari}}]{2020PhRvD.101d4033B}
{Bhagwat}, S., {Forteza}, X.~J., {Pani}, P., \& {Ferrari}, V. 2020, \prd, 101,
  044033, \dodoi{10.1103/PhysRevD.101.044033}

\bibitem[{{Bhagwat} {et~al.}(2018){Bhagwat}, {Okounkova}, {Ballmer}, {Brown},
  {Giesler}, {Scheel}, \& {Teukolsky}}]{2018PhRvD..97j4065B}
{Bhagwat}, S., {Okounkova}, M., {Ballmer}, S.~W., {et~al.} 2018, \prd, 97,
  104065, \dodoi{10.1103/PhysRevD.97.104065}

\bibitem[{{Boyle} {et~al.}(2019){Boyle}, {Hemberger}, {Iozzo}, {Lovelace},
  {Ossokine}, {Pfeiffer}, {Scheel}, {Stein}, {Woodford}, {Zimmerman},
  {Afshari}, {Barkett}, {Blackman}, {Chatziioannou}, {Chu}, {Demos}, {Deppe},
  {Field}, {Fischer}, {Foley}, {Fong}, {Garcia}, {Giesler}, {Hebert}, {Hinder},
  {Katebi}, {Khan}, {Kidder}, {Kumar}, {Kuper}, {Lim}, {Okounkova}, {Ramirez},
  {Rodriguez}, {R{\"u}ter}, {Schmidt}, {Szilagyi}, {Teukolsky}, {Varma}, \&
  {Walker}}]{2019CQGra..36s5006B}
{Boyle}, M., {Hemberger}, D., {Iozzo}, D. A.~B., {et~al.} 2019, Classical and
  Quantum Gravity, 36, 195006, \dodoi{10.1088/1361-6382/ab34e2}

\bibitem[{{Brito} {et~al.}(2018){Brito}, {Buonanno}, \&
  {Raymond}}]{2018PhRvD..98h4038B}
{Brito}, R., {Buonanno}, A., \& {Raymond}, V. 2018, \prd, 98, 084038,
  \dodoi{10.1103/PhysRevD.98.084038}

\bibitem[{{Buchner} {et~al.}(2014){Buchner}, {Georgakakis}, {Nandra}, {Hsu},
  {Rangel}, {Brightman}, {Merloni}, {Salvato}, {Donley}, \&
  {Kocevski}}]{Buchner_AAP2014}
{Buchner}, J., {Georgakakis}, A., {Nandra}, K., {et~al.} 2014, \aap, 564, A125,
  \dodoi{10.1051/0004-6361/201322971}

\bibitem[{Carullo {et~al.}(2019)Carullo, Del~Pozzo, \&
  Veitch}]{PRD_Carullo2019}
Carullo, G., Del~Pozzo, W., \& Veitch, J. 2019, Phys. Rev. D, 99, 123029,
  \dodoi{10.1103/PhysRevD.99.123029}

\bibitem[{{Dreyer} {et~al.}(2004){Dreyer}, {Kelly}, {Krishnan}, {Finn},
  {Garrison}, \& {Lopez-Aleman}}]{2004CQGra..21..787D}
{Dreyer}, O., {Kelly}, B., {Krishnan}, B., {et~al.} 2004, Classical and Quantum
  Gravity, 21, 787, \dodoi{10.1088/0264-9381/21/4/003}

\bibitem[{{Feroz} {et~al.}(2009){Feroz}, {Hobson}, \&
  {Bridges}}]{2009MNRAS.398.1601F}
{Feroz}, F., {Hobson}, M.~P., \& {Bridges}, M. 2009, \mnras, 398, 1601,
  \dodoi{10.1111/j.1365-2966.2009.14548.x}

\bibitem[{{Feroz} {et~al.}(2019){Feroz}, {Hobson}, {Cameron}, \&
  {Pettitt}}]{2019OJAp....2E..10F}
{Feroz}, F., {Hobson}, M.~P., {Cameron}, E., \& {Pettitt}, A.~N. 2019, The Open
  Journal of Astrophysics, 2, 10, \dodoi{10.21105/astro.1306.2144}

\bibitem[{{Finch} \& {Moore}(2021)}]{2021PhRvD.103h4048F}
{Finch}, E., \& {Moore}, C.~J. 2021, \prd, 103, 084048,
  \dodoi{10.1103/PhysRevD.103.084048}

\bibitem[{{Giesler} {et~al.}(2019){Giesler}, {Isi}, {Scheel}, \&
  {Teukolsky}}]{Overtone_PRX_Giesler2019}
{Giesler}, M., {Isi}, M., {Scheel}, M.~A., \& {Teukolsky}, S.~A. 2019, Physical
  Review X, 9, 041060, \dodoi{10.1103/PhysRevX.9.041060}

\bibitem[{Hawking(1972)}]{Hawking:1971vc}
Hawking, S.~W. 1972, Commun. Math. Phys., 25, 152, \dodoi{10.1007/BF01877517}

\bibitem[{{Isi} \& {Farr}(2021)}]{2021arXiv210705609I}
{Isi}, M., \& {Farr}, W.~M. 2021, arXiv e-prints, arXiv:2107.05609.
\newblock \doarXiv{2107.05609}

\bibitem[{{Isi} {et~al.}(2021){Isi}, {Farr}, {Giesler}, {Scheel}, \&
  {Teukolsky}}]{2021PhRvL.127a1103I}
{Isi}, M., {Farr}, W.~M., {Giesler}, M., {Scheel}, M.~A., \& {Teukolsky}, S.~A.
  2021, \prl, 127, 011103, \dodoi{10.1103/PhysRevLett.127.011103}

\bibitem[{Isi {et~al.}(2019)Isi, Giesler, Farr, Scheel, \&
  Teukolsky}]{NoHair_PRL_Isi2019}
Isi, M., Giesler, M., Farr, W.~M., Scheel, M.~A., \& Teukolsky, S.~A. 2019,
  Phys. Rev. Lett., 123, 111102, \dodoi{10.1103/PhysRevLett.123.111102}

\bibitem[{{Kamaretsos} {et~al.}(2012){Kamaretsos}, {Hannam}, {Husa}, \&
  {Sathyaprakash}}]{2012PhRvD..85b4018K}
{Kamaretsos}, I., {Hannam}, M., {Husa}, S., \& {Sathyaprakash}, B.~S. 2012,
  \prd, 85, 024018, \dodoi{10.1103/PhysRevD.85.024018}

\bibitem[{{London} {et~al.}(2014){London}, {Shoemaker}, \&
  {Healy}}]{2014PhRvD..90l4032L}
{London}, L., {Shoemaker}, D., \& {Healy}, J. 2014, \prd, 90, 124032,
  \dodoi{10.1103/PhysRevD.90.124032}

\bibitem[{{Nollert}(1999)}]{1999CQGra..16R.159N}
{Nollert}, H.-P. 1999, Classical and Quantum Gravity, 16, R159,
  \dodoi{10.1088/0264-9381/16/12/201}

\bibitem[{{Pan} {et~al.}(2011){Pan}, {Buonanno}, {Boyle}, {Buchman}, {Kidder},
  {Pfeiffer}, \& {Scheel}}]{2011PhRvD..84l4052P}
{Pan}, Y., {Buonanno}, A., {Boyle}, M., {et~al.} 2011, \prd, 84, 124052,
  \dodoi{10.1103/PhysRevD.84.124052}

\bibitem[{{Press}(1971)}]{GW_APJL_Press1971}
{Press}, W.~H. 1971, \apjl, 170, L105, \dodoi{10.1086/180849}

\bibitem[{Robinson(1975)}]{PhysRevLett.34.905}
Robinson, D.~C. 1975, Phys. Rev. Lett., 34, 905,
  \dodoi{10.1103/PhysRevLett.34.905}

\bibitem[{{Schnittman} {et~al.}(2008){Schnittman}, {Buonanno}, {van Meter},
  {Baker}, {Boggs}, {Centrella}, {Kelly}, \&
  {McWilliams}}]{2008PhRvD..77d4031S}
{Schnittman}, J.~D., {Buonanno}, A., {van Meter}, J.~R., {et~al.} 2008, \prd,
  77, 044031, \dodoi{10.1103/PhysRevD.77.044031}

\bibitem[{{Teukolsky}(1973)}]{QNM_APJ_Teukolsky1973}
{Teukolsky}, S.~A. 1973, \apj, 185, 635, \dodoi{10.1086/152444}

\bibitem[{Thrane {et~al.}(2017)Thrane, Lasky, \& Levin}]{Thrane:2017lqn}
Thrane, E., Lasky, P.~D., \& Levin, Y. 2017, Phys. Rev. D, 96, 102004,
  \dodoi{10.1103/PhysRevD.96.102004}

\bibitem[{{Vishveshwara}(1970)}]{Schw_PRD_Vishveshwara1970}
{Vishveshwara}, C.~V. 1970, \prd, 1, 2870, \dodoi{10.1103/PhysRevD.1.2870}

\end{thebibliography}

\end{document}